# The Ultimate DataFlow for Ultimate SuperComputers-on-a-Chip, for Scientific Computing, Geo Physics, Complex Mathematics, and Information Processing

Veljko Milutinovic, Erfan Sadeqi Azer, Kristy Yoshimoto, Indiana University, Bloomington, Indiana, USA

Gerhard Klimeck, Purdue University, IN, USA

Miljan Djordjevic, Milos Kotlar, Miroslav Bojovic, Bozidar Miladinovic, Nenad Korolija, and Stevan Stankovic, Universty of Belgrade, Serbia

Nenad Filipović, Zoran Babovic, Universty of Kragujevac, Serbia

Miroslav Kosanic, MIT, Cambridge, MA, USA

Akira Tsuda, Harvard University, Cambridge, Massachusetts, USA

Mateo Valero, BSC, Barcelona, Spain

Massimo De Santo, University of Salerno, Fisciano, Italy

Erich Neuhold, UNIWIE and TUWIEN, Vienna, Austria

Jelena Skoručak, University of Zurich and ETH, Switzerland

Laura Dipietro, Highland Instruments, Cambridge, MA, USA

Ivan Ratkovic, Esperanto Technologies, Belgrade, Serbia and San Francisco, California, USA

This article starts from the assumption that near future 100BillionTransistor (100BT) SuperComputers-on-a-Chip will include N big multi-core processors, 1000N small many-core processors, an ASIC TPU-like fixed-structure systolic array accelerator for the most frequently used Machine Learning algorithms needed in bandwidth-bound applications and an FPGA flexible-structure re-programmable accelerator for less frequently used Machine Learning algorithms needed in latency-critical applications. Of course, appropriate interfaces to memory and standard I/O, as well as to the Internet and intranet, plus big data collectors and various external accelerators are absolutely necessary, as depicted in the attached figure. The future SuperComputers-on-a-chip should include effective interfaces to specific external accelerators based on Quantum, Optical, Molecular, and Biological paradigms (programmable using programming models referred to here as Energy Flow), but these issues are outside the scope of this article. Big data collectors could be storage-based (like IoT, Internet of Things) or sensor-based (like WSN, Wireless Sensor Networks), both mass-programmable using programming models referred to here as Diffusion Flow. The number of processors in Figure 1 could be additionally increased if appropriate techniques are used, like cache injection and cache splitting [15, 16]. Finally, a higher speed could be achieved if some more advanced technology is used, like GaAs [13,14]. This group has an interest in applying the described hundred billion transistor architecture for artificial intelligence and its system support [21,22]. Figure 1 is further explained with data in Table 1, which is aimed at 100BT chip, and could serve as an introduction for future efforts related to 1TrillionTransistor (1TT), like [36]. The architecture that we consider here is meant for applications not reachable by quantum computers, of Qubit type [40] or photonic type [41].

This article also advocates the need for two new programming models that could be used to program two newly emerging computing paradigms mentioned above:

1. The Diffusion Flow Programming Paradigm

    The I/O of the chip under consideration is linked
    to 4 basic sources/sinks of Big Data:
    WSN, IoT, Internet, intranet.
    Data could be pre-processed before input into the chip!

    This programming paradigm could be used for data processing
    within a WSN (Wireless Sensor Network),
    before the refined data are passed via I/O
    to the sCC (SuperComputer-on-a-Chip) described in this article.

    If various nodes on an IoT (Internet of Things)
    could communicate with each other,
    than the pre-processing of the IoT data
    could also be done within the IoT, before being passed on via I/O.

    Before input into the sCC (SuperComputer-on-a-Chip),
    data may have to be collected from the Internet,
    and filtered or modified in some proper method,
    so a prorer programming paradigm is needed for that purpose, too!

    Intranet data coming via a stream may also need pre-processing
    before reaching the central structure elaborated in this article,
    and these often propagate like a diffusion,
    which determines the name for this new programming model.

    This means that pre-processing would occur as the data
    propagate through the system, meaning that the basic notions
    of approximate Turing computing have to be used,
    since the processing power of WSN and IoT processors is limited,
    while the data coming tom Internet and intranet, could be insufficient!

2. The EnergyFlow Programming Paradigm

    External accelerators of the chip, in essence,
    do some energy transformations while data is being processed,
    for algoritms not suitable for ControlFlow or DataFlow.
    The programming model should be aware of the related energy flows.
    It would be ideal if the same programming model,
    based on energy awareness could be elaborated
    for Biology-based computing, for Molecular-based computing,
    for Opto-based computing, and for Quantum-mechanical computing.

Of course, the language constructs should reflect the needs
of the major applications that the sCC would be intended for:
Scientific Computing (in Physics, Chemistry, Biology, and Genomics),
Geo Physics (with emphasis on Weather Forecast, Oil and Gas),
Complex Mathematics (Tensor Calculus and Artificial Intelligence),
and Information Processing using Big Data.
The language extensions of popular programming languages
should include two basic domains:

(A) Built-in classes to make the programming easier,
(B) Built-in data types that reflect the way in which
the data are propagated through the computing infrastructure!

Of all existing and popular languages,
the best candidate to use as the basis for these two extensions
is the language with the smallest Kolmogorov Complexity!

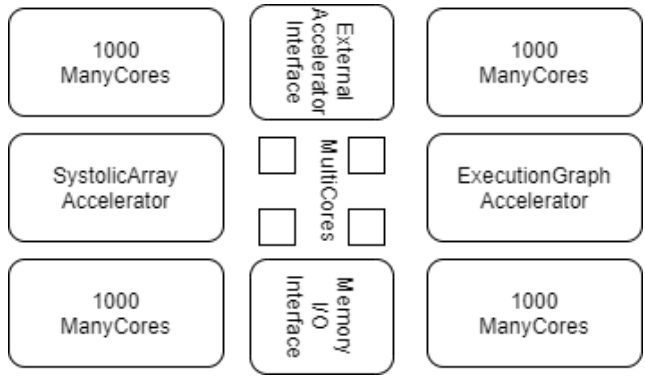

**Figure 1:** Generic structure of a future SuperComputer-on-a-Chip with 100 Billion Transistors

| Chip Hardware Type | Estimated Transistor Count |
|---|---|
| One ManyCore with Memory | 3.29 million |
| 4000 ManyCores with Memory | 11 800 million [17] |
| One MultiCore with Memory | 1 billion [18] |
| 4 MultiCore with Memory | 4 billion |
| One Systolic Array | <1 billion [19] |
| One Reprogrammable Ultimate Dataflow | <69 billion [20] |
| Interface to I/O with external Memory | <100 million |
| Interface to External Accelerators | <100 million |
| TOTAL | <100 billion |

**Table 1.** Basically, current efforts include about 30 billion transistors on a chip, and this article advocates that, for future 100 billion transistor chip, the most effective resources to include are those based on the dataflow principle. For some important applications, such resources bring significant speedups, that would fully justify the incorporation of additional 70 billion transistors. The speedups could be, in reality, from about 10x to about 100x, and the explanations follow in the rest of this article.

Some current efforts aimed at putting about 30-60 billion transistors onto a chip that includes CPU-like and/or GPU-like engines, are summarized in Table 2, together with a recent 1TT effort.

| Company | Product name | Country | City | Number of CPU cores | Number of GPU cores | CPU clock rate | Launched |
|---|---|---|---|---|---|---|---|
| Alibaba | XT910 | China | Hangzhou | 1/2/4 per cluster | N/A | $2.0 - 2.5$ GHz | 2020 |
| RISC-V | Micro Magic RISC-V Core [39] | US | San Francisco | 1 | N/A | $4.25 - 5.19$ GHz | 2020 |
| SiFive | FU740 RISC-V SoC [25] | US | San Francisco | 4 | 1 | $1.4 - 1.5$ GHz | 2020 |
| AMD | AMD Ryzen™ 5 3400G with Radeon™ RX Vega 11 Graphics [24] | US | Santa Clara | 4 | 11 | $3.7 - 4.2$ GHz | 2019 |
| Nvidia | Tegra Xavier [28] | US | Santa Clara | 8 | 512 CUDA | N/A | 2019 |
| Esperanto Technologies | N/A | US | Mountain View | 16 ET-Maxion cores | 4096 ET-Minion cores + ET-Graphics [26,27] | 2+GHz | 2018 |
| Intel | Intel Sandy Bridge [23] | US | Santa Clara | 1-4 (4-6 Extreme, 2-8 Xeon) | 6 | $1.60 - 3.60$ GHz | 2011 |
| Moscow Center of SPARC Technologies (MCST) | Elbrus-2S+ (1891ВМ7Я) [29] | Russia | Moscow | 2 Elbrus 2000 cores | 4 DSP Elcore-09 cores | $300 - 800$ MHz | 2011 |

**Table 2.** Current chips paving the way to an Ultimate SuperComputer-on-a-Chip

Since the first three structures (multi-cores, many-cores, and TPU) are well elaborated in the open literature, this article focuses only on the fourth type of architecture, and elaborates on an idea referred to as the Ultimate DataFlow, that offers specific advantages, but requires a more advanced technology, other than today's FPGAs.

In addition, some of the most effective power reduction techniques are not applicable to FPGAs, which is another reason that creates motivation for research leading to new approaches for mapping of algorithms onto reconfigurable architectures. Consequently, the novel approach, referred to as Ultimate DataFlow, is described next. As far as IO and interfaces to external accelerators, a number of specific issues has to be kept in mind. [52]

**Introduction to Ultimate Data Flow**

The architectures like Google TPU are extremely effective for the most frequent Tensor Calculus and related algorithms to which they are tuned, but these algorithms, in many important applications, burn only about 50% of the run time. In these applications, the other about 50% of run time gets burned by a huge number of other algorithms, so their architectural support requires a lot more flexible and fully reconfigurable architecture.

This article sheds light on the newly proposed concept, Ultimate DataFlow for BigData, offering flexibility and reconfigurability in DeepAnalytics (DA) and MachineLearning (ML).

Some of the problems in DA and ML are bandwidth-bound, while the others are latency bound. The bandwidth-bound problems could, for many applications, be solved successfully using the FPGA-based DataFlow systems. The most critical latency-bound problems need a different in-memory computing technology. The Ultimate DataFlow implies elements of internal analog processing, which brings potentials, that are first presented, and then explained through an adequate elaboration.

**Potentials of Ultimate DataFlow**

The Ultimate DataFlow approach offers an effective solution for latency-bound problems, with the following improvement potentials over the FPGA-based solutions:

  (A) Up to about 2000 in speed up,
  (B) Up to about 200 in transistor count,
  (C) Up to about 80 (20x4) in power reduction, and
  (D) Up to about 2 in data precision.

With the above in mind, this position article covers the issues related to the potentials of the concept, using the programming model utilized in numerous FPGA-based DataFlow engines.

The existing DataFlow approaches are still far away from the ideal Ultimate DataFlow, but for specific Machine Learning and BigData applications, they still do achieve considerable speedups over ControlFlow machines, especially for some specific BigData problems (potentials are illustrated in Figure 2). Consequently, the drive for new technology-supported architectural solutions is not too strong these days. However, for research missions, like those around Mubadala and IMEC, or Esperanto and IPSI, innovative solutions are badly needed.

What is good, however, about the existing DataFlow approaches, is that their DataFlow programming models are directly applicable to the case of the Ultimate DataFlow, so the continuity of existing experiences and the already developed software products could be maintained and improved.

For precision, the ratio 2x was quoted above, since the approach could benefit from approximate computing, due to its data format flexibility, as explained later. It is well suited also for *bloat16*, a possible new standard for tensor applications.

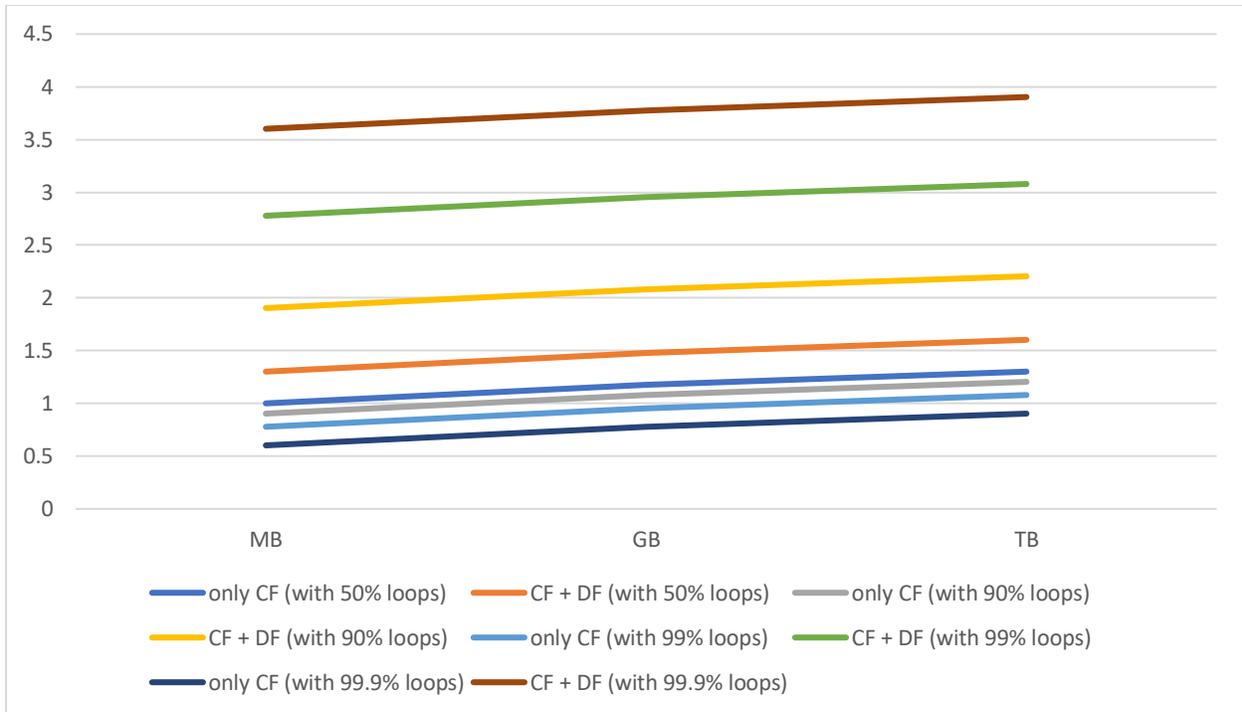

**Figure 2:** Potentials of the ultimate dataflow add-on for various dataflow/conrolflow system speedups, assuming the theoretical case, when time consumed by loops have the zero execution time, after migration to the ultimate dataflow accelerator on the chip. X axis = data. Y axis = logarithmic normalized speedup.

**Elaboration of Potentials of Ultimate DataFlow**

For power, the ratio 80x was quoted, for two reasons: First, ControlFlow machines like Intel or NVidia operate on up to 4GHz or even higher frequencies, while the current FPGAs operate on about 200MHz, which makes the ratio of about 20x, when it comes to dissipation. Second, an additional 4x one could get from operating at the 2x lower voltage. Both factors together result in the total improvement ratio of 80x.

For transistor count, the ratio of 200x was quoted, for the following reason: If one looks up the Intel microprocessor floor plan, one finds out that only about 0.5% of the area is dedicated to Arithmetic and Logic, making the above quoted 1/200x ratio. The UROC (Undergraduate Research) of Kristy Yoshimoto proves that the described dataflow add-on to the one hundred billion transistor chip is a lot more usefull for arithmetic than for logic.

For speedup, the frequently quoted numbers are: (a) 20x as the lowest number on speedup in recent publications of the authors of this article, (b) 200x as the highest number ever reported by the same authors, and (c) 2000x was quoted for the reason, that has nothing to do with existing DataFlow implementations, but has a lot to do with Ultimate DataFlow, as elaborated later; (d) even 20000x could be hoped for some applications, as explained next.

In Ultimate DataFlow, the speedup depends predominantly on the contribution of loops to the overall execution time:

- If loops contribute with more than 99.95% to the overall run time, then one can hope for a speedup of 2000x.

- If one looks up some of the applications on the list of current DataFlow successes in Machine Learning for BigData, one finds out that in many cases the contribution of loops was well over 99.995%, which is why the potentials of Ultimate DataFlow could reach even 20000x.

**Explanations of the Ultimate DataFlow Concept**

The Ultimate DataFlow, as a concept, is built on the following two premises (each one with 4 sub-premises):

1. *Compiler does the following:*

   a) Separates effectively spatial and temporal data, to satisfy the requirements of the Nobel Laureate Ilya Prigogine, since that action lowers the entropy of a computer system, meaning that the rest of the compiler could do a much better optimization job (lower entropy brings more order into the optimization process and consequently better optimization opportunities).
   b) Maps the execution graph in the way that makes sure that edges are of the minimal length, which brings consistency with the observations of Nobel Laureate Richard Feynman, related to trade-offs between speed and power.
   c) Enables one to go to a lower precision, for what is not of ultimate importance, and consequently to save on resources, that could be reinvested into what *is* of ultimate importance, following the approximate computing wisdom of Nobel Laureate Daniel Kahneman.
   d) Enables one to trade between latency and precision, which, in latency-tolerant applications, brings more precision with less resources, and in latency-intolerant applications, brings less latency, in exchange for a lower precision, thus following the wisdom of Nobel Laureate Tom Hunt, and analogies with his findings related to birth, life, reproduction, and death of cells.

None of the FPGA-based dataflow compilers, as far as we know, does any of the above.

2. *Hardware consists of the following:*

   a) An analog DataPath of the honeycomb structure, to which one could effectively map the execution graphs corresponding to loops. Analog functional units could leverage low-precision computation.
   b) A DataPath clocked at a much lower frequency, and hopefully not clocked at all, if the analog path is not unacceptably long, so it is literally the voltage difference between input and output, that moves data through the execution graph.
   c) A digital memory is on the side of the DataPath, so that computing parameters could be kept non-volatile, and temporary results could be stored more effectively.
   d) The I/O connecting the host and the dataflow is much faster.

Unfortunately, FPGAs offer none of the above today! Consequently, the FPGA technology is today only the least bad solution on the road to the ultimate goal!

In conclusion, the benefits of the Ultimate DataFlow approach will become fully achievable only once the semiconductor and the compiler technologies become capable of supporting the above specified two sets of requirements. References leading to the above conclusion are spreading four decades of the research of one of the co-authors [1, 2, 3, 4, 5, 6, 7, 8, 9, 10]. The future is in in-memory analog AI accelerators, as explained in the recent effort of [11]. Another viewpoint of the related issues could be found in [12].

**Potential Applications**

The major application domains, conditionally speaking, belong to the following domains: Health in the widest sense, Genomics, GeoPhysics, Energy, Finances and the processing of Video Data in the widest sense. Of course, collection of big data via sensors and from IoT, as well as the application of machine learning and artificial intelligence in general, are applied.

In the domain of scientific computing, a graduate student from Indiana University has worked on two projects involving the Maxeler dataflow system. In the first project, a randomized embarrassingly parallel algorithm has been developed to estimate the value of π. This algorithm takes advantage of the FPGA-based parallel processor to draw an enormous number of samples in parallel, to aggregate, and to estimate the target value. This is in contrast to other approaches, like Taylor series-based ones, that are inherently sequential.

Accurate approximations of π have essential applications in numerical analysis. In the second project, an edge detection algorithm has been implemented with many applications in the graphic processing field. The used convolution-based approach uses kernel computations in small neighborhoods and fits the dataflow paradigm perfectly.

An innovative AI healthcare system under development, needing high acceleration of complex algorithms, is described in [31]. Advances in deep learning, or data-driven AI, mean that algorithms can generate layers of abstract features that enable computers to recognize complicated concepts (such as a diagnosis in healthcare systems) by building on simpler ones that are accessible in the data. Data-driven AI requires big data and very substantial computing power to reach adequate performance levels. Such algorithms have been around for some time now, but the recent expansion of datasets and computational resources have enabled a series of breakthrough improvements that could now be applied to augment healthcare provision. The number of data sources in healthcare services has grown rapidly as a result of widespread use of mobile, wearable sensors technologies and implants, which has flooded healthcare area with a huge amount of data and changed our understanding of human biology and of how medicines work, enabling personalized and real-time treatment for all. Examples of big data in sports medicine needing acceleration could be found in [32,33,34,35,53,54,55,56].

Of course, all this computing power should be peaceable onto a single chip. Furthermore, wearable heath technologies are becoming increasingly popular in various fields of medicine, both as an aid in clinical diagnostics, and as consumer gadgets – for example, EKG monitoring while running, or sleep monitoring at home. AI technologies are opening new doors to large-scale data processing in medicine which is of considerable importance for both research and clinical applications, as well as for allowing automatic or semi-automatic detection of certain patterns in the patient data recordings.

Some examples of potential applications of AI-based semi-automatic detection (detected by the algorithm, and checked by the clinician) are detections of various heat arrhythmias in the electrocardiogram (EKG) recordings 1,2, identification of sleep episodes in the electroencephalogram (EEG) data as an aid to diagnose sleep disorders 3,4, detection of pneumonia in computer tomography (CT) images in order to diagnose COVID19 [57], etc. All these algorithms could profit from higher computational power, such as dataflow, as the increase in the computational power was very important for the widespread application of AI in the past years. Another important benefit of AI and higher computational power is personalization in the medical field. With the increase in the computational power, there will be no necessity to apply a single algorithm to all individuals, but the algorithm could be re-trained to fit individual needs. For example, in case of sleep consumer wearables, a pre-trained algorithm could be re-trained locally on a powerful chip integrated in the consumer device, which would result in the personalization of the algorithm to fit individual data. Of course, all this computing power should be placeable onto a single chip.

Genomics research is now entering the age of millions of available genomes. Possible solutions to bigger data loads of future genomics projects involve cloud and dataflow computing. Data generated through DNA sequencing is rapidly becoming big enough for this problem to take center stage as the biggest big data problem of today. The amount of data created this way will soon surpass digital information available in astrophysics, particle physics and even on some of the largest Internet sites. [37]

Seismic data is growing exponentially. Earth's inner layers of different rock qualities reflect energy, which is recorded and stored as terabytes and petabytes of data. 2D and 3D models of the inner layers can be generated using this data, and heavy computation required for this process can be vastly improved through the use of dataflow machines. [38]

Power system security assessment is an extremely computationally demanding task. Size of the problem becomes intractable for systems exceeding the size of a few buses. To address this issue efficiently, machine learning techniques can be employed. Convolutional neural networks (CNN) may be applied to assess the state of the power system when used with the images formed as state snapshots [49], but they are burdened with computationally intensive training. As was shown in [47] and [48] a significant time improvement can be gained through Dataflow paradigm.

The focus of modern financial algorithms is on speed, accuracy and precision. Due to rapidly changing conditions, their performance in times of economic disturbances, as they are further enhanced, shouldn't be degraded. It is a usual case with the complex simulations that time execution becomes a bottleneck with the increase of the number of trials or the number of steps. Such an example would be a simulation of Heston processes [50], namely the procedure called "Exact scheme", a method developed by Broadie and Kaya [51] to exactly deduce volatility of an asset. We advocate that implementing this simulation using DF would drastically reduce its execution time.

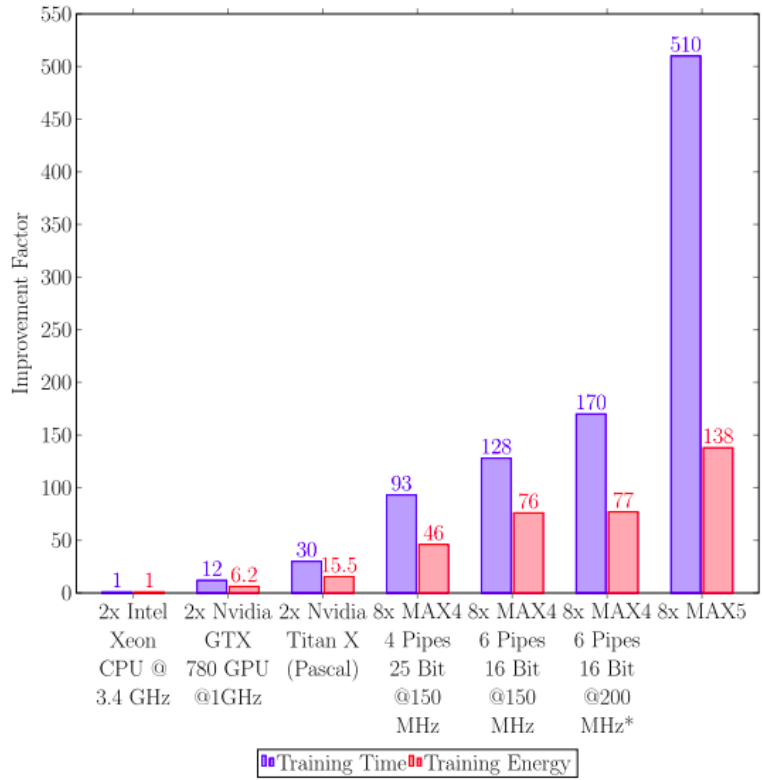

**Figure 3:** CNN training speedup, result taken from presentation [47] which is part of the paper [48]

The new Ultimate DataFlow technique seems useful for two areas where Big Data are involved:

(i) The AI approach of segmentation in lung image studies. Particularly, we have to train the system from huge data. Perhaps, in this process, "Ultimate DataFlow" technique would be very useful.

(ii) Another possibility are the alveolar deposition studies. As it is known, there are several hundred million alveoli in the lungs. With the "Ultimate DataFlow" technique, personalized approaches could be possible.

In robotics, supercomputing has so far been little explored and limited to optimization of complex robotics tasks. However, development of supercomputers on-a-chip, coupled with machine learning and other AI algorithms, promise to significantly advance the field. Autonomous, mobile robots can particularly benefit from on-board supercomputer systems for data processing. For example, adding supercomputers on board of aerial mobile robots or drones can allow complex data processing and delivery of immediate results (see, for example, [42]); in multi-robot systems (such as cooperative or swarm robots) control, real-time complex data processing allowed by supercomputers on-a-chip can allow implementation of increasingly complex control strategies. As autonomous, mobile robots are increasingly being used, e.g., in areas that are remote or too dangerous to reach for human workers, these enhanced capabilities are becoming increasingly necessary. Another area of robotics where supercomputers on-a-chip are likely to make an impact is medical robots that rely on intensive processing (e.g., for modeling/simulations of surgical interventions) and need to be deployed in non-conventional medical environments. Key requirements for all of the above applications are low power consumption, small size, scalability (especially for applications involving multi-robot systems), and flexible dataflow management.

In the past years, there has been a trend to develop neuromorphic supercomputers that mimic the structure and our current understanding of the functioning of the nervous system (see, for example, Neurogrid developed at Stanford University). Their implementation on chips is expected to change AI and find application in many fields including robotics. This research trend is not surprising, given the proliferation of neuroscience discoveries on information processing principles implemented by the brain and the fact that biology has often served as a basis of inspiration for design of new technologies. Furthermore, the brain has very desirable engineering features: it is extremely energy efficient, has a compact design, high fault tolerance, and fast processing speed. These features are the result of a combination of factors, such as its architectural design, dataflow management strategy, and evolution. As this research field is still relatively young, a number of research questions remain unexplored. For example, it is unclear which level of replication of the nervous system is needed for which applications or which improvements can be achieved (e.g., in dataflow management by implementing mechanisms similar to those that are used by the brain to correctly function over time (such as pruning of electrical connections or re-wiring after damage)). Implementation of hybrid design strategies from classical computer engineering and that are neuroscience-inspired have a high potential to advance development of this type of supercomputer.

A possible application of supercomputers on a chip in robotics is in swarm robots control and the main requirement is scalability with flexible data management. The dataflow management, in general, is thought-provoking, and we wonder to what extent strategies for efficiency that are implemented by the brain, e.g. pruning of electrical connections, are implementable in dataflow management.

The brain is the most powerful supercomputer, it takes little space, and has a supercompact design. Computer designers should look into its functioning and get some ideas! This a possible topic of interest for future research!

Through aggressive semiconductor device down-scaling over the past 50+ years we have today, in 2021, industrial, massively produced transistors guide the electron flow through critical dimensions of a few nanometers. At these atomic dimensions an approach to understand, model, design, and optimize devices must be very different from classical large-scale semiconductor devices that are described in yesterday's textbooks and commercial software. The physical structure must be based on and described with individual atoms which at no place imposes infinitely periodic geometries, but comprehends finite spatial extensions. Within the physical structure the electronic structure must be comprehended fully quantum mechanically in a spatially varying finite device. This quantum mechanical system is *not* closed but explicitly open, which alters the quantum mechanical states in the central device system. Supriyo Datta introduced the Non-Equilibrium Green Function formalism (NEGF) into the electrical engineering community [43]. Klimeck's teams utilized NEGF combined with empirical tight-binding and novel computational and user interface approaches to build the Nanoelectronic Modeling (NEMO) tool suites [44]. NEMO tools have been used to impact actual transistor designs such as the rotated substrate at Texas Instruments 18 years ago and are being licensed and adopted in industry in the past 10 years. The rotated substrate (crystal anisotropy) and the strain engineering are examples where the quantum mechanical nature of the electrons at the nanometer scale can be used to improve device performance. Yet typically most device engineers think of the quantum mechanical effects as detrimental that need to be overcome.

In contrast to classical transistor devices research has been conducted for the past 40+ years into a very different class of devices: Quantum Devices. Here the quantum mechanical nature of the electrons and quantum states is to be explicitly utilized to ultimately process information. In this field devices are typically considered to be closed systems in a desired isolation from their surrounding environment. Even systems of multiple qubits or optical quantum dots are typically treated as just perturbatively coupled to their environment. The empirical tight binding methodologies in NEMO are applicable and have been finely tuned to such closed quantum mechanical systems. However, the computation of quantum states in closed (Hermitian) systems versus open (non-Hermitian) systems requires for practical device applications very different computational algorithms and data structures.

The modeling of electronic structure in closed systems typically boils down to the computation of interior spectrum eigenvalues and eigenvectors in system sizes of up to $10^9$ x $10^9$. For example electron states in a single P impurity have a very shallow binding energy below the conduction band and the wavefunction spreads out to a space of 30x30x30 $nm^3$ corresponding to about 1.7 million atoms. A strained InAs on GaAs quantum dot of lateral dimensions 10x10x5 $nm^3$ needs a similar domain to account for wavefunction penetration into the buffer layer. A very good choice of tight binding basis set is typically the nearest-neighbor sp3d5s* which with explicit spin accounts for 20 orbitals per atom. With 1.7 $10^6$ atoms we get to a very sparse system matrix of 3.4x$10^6$. The matrix contains 20x20 dense blocks on the diagonal and 4 such additional blocks in each row and column to connect the 4 neighboring atoms.

Reasonable ordering makes the matrix diagonally dominant. NEMO uses a custom Lanczos algorithm [45] to compute eigenvalues and vectors that now basically multiplies the sparse system matrix repeatedly with a trial and refinement vector. In the early days of NEMO3D (1998-2003) the mere storage of the system matrix was an memory issue and the matrix elements had to be computed on the fly for large systems.

Later we found it very beneficial to customize the low-lying small block matrix multiplications to available architectures like Intel SSE instruction sets. We tried but failed to map the problem efficiently into GPU based architectures (2007-2009) and did not achieve any real efficiency gains.

Open system simulations utilizing NEGF require significantly different algorithms. The system matrix is based on the one described above with in critical addition. The atoms that are "attached" to one specific open boundary surface are becoming all connected and form a large dense block, typically on the top left and the bottom right of the system matrix. Each dense block typically represents a slice of the device in the transport direction. The most effective algorithm to solve this system reliably is the Recursive Green Function (RGF) algorithm [46] which requires the inversion of these device layer blocks and a few matrix-matrix multiplications per layer. Our research team spent significant efforts 2007-2014 to customize such algorithms to Intel Phi and GPUs but were ultimately not able to obtain any significant simulation time reductions. These efforts included low level CUDA programming as well as higher level MAGMA coding or simple off-loading. We abandoned the efforts to pursue further code optimizations based on GPUs.

In summary did find that low level customization of low-level block-matrix-vector multiplications using the Intel SSE instruction set to be very useful and not particularly intrusive to the overall code structure. However, we have spent significant sponsor-funded efforts on customizing nano-electronic algorithms to Intel-Phi as well as GPU-based hardware and found that we were not able to achieve any significant speedup on these platforms. What was even more disheartening was that these customizations implied significant code modifications that were not just at the lowest level computational unit, but they required some architectural interventions that mater the maintenance of the overall code base more difficult.

Our research team builds codes to the solution of specific modeling problems. As such, the physics solution is of utmost importance. Computational approaches and hardware that require significant code re-writes or re-architecting are at odds even with midsize research groups that need to maintain a certain continuity in s/w development. As such new architectures can only be adopted by us within our limited s/w development resources.

Utilization of the elaborated chip technology has implications even in fields like high-resolution photography:

1) PRICE: Less expensive lenses bring the equipment cost down but require sophisticated computing: The 6 aps-c cameras (~6x330$=1980$ + one small sensor) versus one medium format camera (~4500$).

2) SPEED: Multi-higher-speed image processing and performance cameras become an affordable reality (slightly lower speed than on aps-c cameras versus medium format camera speed).

3) QUALITY: Higher image quality (higher resolution, higher dynamic range, better ISO performance ...): Total six sensor size of about 5cmx5cm or more versus one medium format of about 3.3cm x 4.7cm.

4) USABILLITY: Already existing lenses for medium format cameras could be used (so we have space for the sensor to be up to 6x6cm in size).

A fascinating field of application of the Ultimate Dataflow approach's potential is in supporting context-driven cultural tourism services.

As a part of the activities of the University of Salerno and the High Technology District for Cultural Heritage of the Campania Region, we have been able to experiment with the creation of mobile-oriented environments that allow tourists interested in building a personalised itinerary to discover the cultural heritage, in particular at the Urban Archaeological Park of Piazza Municipio in Naples, Italy. The application is based on the following assumptions:

- the ability to analyse in real-time the flow of data coming from the IoT sensors placed in the 120 Points of Interest defined in the area subject of the intervention, making available with continuity statistics related to environmental data and the monitoring parameters of the heritage components located there

- the ability to analyse in real-time the video streams coming from the more than 200 cameras located on the sites

- the ability to propose customised itineraries, dynamically adapted to the available context information, based on a match between the user's profile and preferences and the available tourist-cultural offer.

As it is evident from the above, we are in the presence of a Big Data context (given the massive amount of information coming from sensors) where AI techniques for natural language interaction with the user are a fundamental prerequisite for achieving an adequate level of user satisfaction. Therefore, the use of new technology seems to be a natural solution to the demanding requirements of computational power and adequate flexibility that the described scenario requires.

Ultimate DataFlow approach can be applied for the modelling of drug delivery to cancerous cells. The use of drug delivery systems has already revolutionised the areas of cancer prevention and pain management related to classical cancer chemotherapy. Furthermore, the encapsulation of the drug in a microreservoir can improve drug solubility and stability as well as reduce drug resistance in some human cancers. Regarding computational methods, it can be stated that there is no general concept for mass transport within vessels (large and capillary blood vessels and lymph), extracellular space and cell interior. Described problems fall under multiscale modelling and our aim is to use either continuum/discrete modelling as well as deterministic/stochastic approach in order to provide a full model for simulation of the therapeutic nanosystems delivery to the cancerous lesions. The basis for a general drug delivery to cancerous cells relies on Finite Element Modelling - the use of different elements 1D (truss, beam), 2D (axisymmetric, membrane, plate) and 3D elements (tetrahedra, brick with/without midside nodes) including recently developed Composite Smeared Finite Element (CSFE) [58], with improvements of accuracy [59], further enhanced to include lymphatic system [60], and generalized as a multiscale element which includes cell interior with organelles [61].

Also some application has been recently published for COVID-19 [62]. Some possible application of Ultimate DataFlow approach for large scale cardiac modeling is 3D deformable body which represents the left and right ventricle. Blood flow is modeled during filling phase by applying the fluid-solid interaction method. The ventricle wall is modeled by 3D brick 8-node solid elements, with fibers that have three-dimensional direction. The Navier-Stokes equations are solved using the ALE formulation for fluid with large displacements of the boundary. In addition, a remeshing procedure will be employed for the fluid domain in accordance with the motion of the ventricle wall. The ventricle wall model will be simulated by muscle material model. Muscle fiber orientation is defined by direction vector in 3D prescribed through input data. The outlet blood pressure is used as the boundary condition.

At the same time, the wall muscle fibers are activated according to the activation function taken from specific patient measurement.

## Experiences in Education and Research

About 4000 students world-wide have used the dataflow machine at the Mathematical Institute of the Serbian Academy of Sciences (https://maxeler.mi.sanu.ac.rs/), and these students come from universities like: MIT, Harvard, Princeton, Yale, Columbia, NYU, Purdue, University of Indiana in Bloomington, University of Michigan in Ann Arbor, Ohio State, Georgia Tech, CMU, FIU, FAU, etc (in the USA), ETH, EPFL (in Switzerland), UNIWIE, TUWIEN (in Austria), Karlsruhe, Heidelberg (in Germany), Manchester, Bristol, Cambridge, Oxford (in England), and, of course, from the leading schools of Belgrade: ETF, MATF, FON, FFH. They attended the hands-on workshops of classes for one, two, three, or six credits.

As far as research efforts with students, they were asked to compare a real ControlFlow Multicore, a real Controlflow ManyCore, a real FPGA-based DataFlow, and a theorethical Ultimate DataFlow machine based on an analog Sea-of-Gates architecture. Esspecially intensive was the students-oriented research effort at the University of Indiana, since early 2016, through courses on DataFlow SuperComputing (for BigData) and Software Engineering Management (with Creativity Methods), plus through the undergraduate research effort called UROC.

Two UROC students have contributed significantly to programming that demonstrates the potentials of Ultimate DataFlow. Students in Siena, Salerno, Barcelona, and Valencia contributed to the development of related concepts.

The Belgrade University graduate and undergraduate students helped determine the best distribution of transistors over resources, for a possible effort based on a 100 billion transistor chip. These efforts were oriented to Tensor Calculus [30].

February 8, 2021

## Acknowledgements:

The authors are thankful to Lars Zetterberg of KTH, Henry Markram of EPFL, Roberto Giorgi of the University of Siena, and Anton Kos of the University of Ljubljana, for their eyes opening discussions related to the topic of ICT.